\documentstyle[preprint,aps]{revtex}
\begin{document}
\tighten
\draft
\title{Non-equilibrium perturbation theory for complex scalar fields}
\author{I.D. Lawrie\thanks{email: i.d.lawrie@leeds.ac.uk} and D.B.
McKernan\thanks{email: phybm@phys-irc.novell.leeds.ac.uk}}
\address{Department of Physics and Astronomy, University of Leeds,
Leeds LS2 9JT, England}
%\date{\today}
\maketitle
\begin{abstract}
Real-time perturbation theory is formulated for complex scalar fields
away from thermal equilibrium in such a way that dissipative effects
arising from the absorptive parts of loop diagrams are approximately
resummed into the unperturbed propagators. Low order calculations of
physical quantities then involve quasiparticle occupation numbers
which evolve with the changing state of the field system, in contrast
to standard perturbation theory, where these occupation numbers are
frozen at their initial values. The evolution equation of the
occupation numbers can be cast approximately in the form of a
Boltzmann equation.  Particular attention is given to the effects of a
non-zero chemical potential, and it is found that the thermal masses
and decay widths of quasiparticle modes are different for particles
and antiparticles.   
\end{abstract}
\pacs{11.10.Wx; 05.30.-d; 98.80.C}
%%%%%%%%%%%%%%%%%%%%%%%%%%%%%%%%%%%%%%%%%%%%%%%
% SECTION 1
%%%%%%%%%%%%%%%%%%%%%%%%%%%%%%%%%%%%%%%%%%%%%%%
\section{INTRODUCTION}
Many problems in physics require an understanding of the
non-equilibrium properties of highly excited states of a quantum field
theory. The principal motivation underlying the present work is to
understand the nature of phase transitions in the very early universe,
in particular those that might give rise to inflation
\cite{guth:prd,linde:pl,albrecht:prl,abbott}.  Most discussions of
inflation assume that the non-equilibrium state of the relevant system
of quantum fields can adequately be characterised by expectation
values governed by a classical potential.  However, calculations which
attempt to go beyond this classical approximation indicate that it may
be seriously inaccurate (see e.g. \cite{boyan1,boyan2,boyan3,boyan4}).
Non-equilibrium effects are also likely to be important in the context
of baryon-number violation at the electroweak scale
\cite{kuzmin,cohen} and in the study of quark-gluon plasmas formed in
heavy-ion collisions \cite{mueller}, as well as in many aspects of
condensed matter physics.

Our own long-term programme, set out in \cite{idl:npb}, has as its
goal a complete description within perturbation theory of the dynamics
of symmetry-breaking phase transitions. An essential requirement is to
formulate perturbation theory in such a way that the evolving state of
the non-equilibrium quantum fields is taken adequately into account in
the low orders of the perturbative expansion which are likely to be
tractable.  The standard closed-time-path formalism
\cite{schwinger,mahan,bakshi,keldysh,chou,landsmann} which deals with
real-time phenomena in thermal equilibrium was generalised by Semenoff
and Weiss \cite{semenoff} to apply to a real scalar field with a
time-dependent action, such as might arise in a Robertson-Walker
spacetime.  However, the propagators entering their Feynman rules
depend on the state of the field essentially through single-particle
occupation numbers which are fixed at their initial values, and
therefore do not properly reflect the evolution of the state.

In earlier work \cite{idl:prd,idl:jpa,idl:cjp} we have shown, for real
scalar fields how this difficulty may be overcome. By adding a
suitable counterterm (which we call the ``dissipative'' counterterm)
to the quadratic part of the action, and subtracting it from the
interaction, we obtain a lowest-order theory in which the absorptive
parts of loop contributions to the 2-point functions are partly
resummed into the unperturbed propagators. As a result, these
propagators contain quasiparticle occupation numbers which evolve with
time in the expected way. Interestingly, the form of the counterterm
is rather similar to the action encountered by Hu and various
collaborators (see e.g. \cite{hu1,calzetta1,hu2} and references given
in these papers). These authors use the Feynman-Vernon influence
functional technique to obtain the effective action for a quantum
field in the presence of environmental variables (which can be thought
of as a heat bath) when the latter are integrated out.  Our procedure
can be thought of as a self-consistent treatment of a field which
provides its own heat bath, though none of the modes of this field are
integrated out. A similar structure also arises in non-equilibrium
formulations of thermo-field dynamics (see e.g. \cite{hardman}) but
this formalism turns out to equivalent to a version of quantum
statistical mechanics restricted to Gaussian initial density matrices
\cite{idl:jpa2}

In dealing with a realistic model (such as the standard model or a
grand unified theory), one naturally meets complex scalar, spinor and
gauge fields and it is essential to extend the non-equilibrium
formalism to encompass fields of these kinds. (Real-time perturbation
theory for scalar, spinor and gauge fields in equilibrium is
discussed, for example, by Kobes {\it et al} \cite{kobes}.) The
purpose of the present work is to extend the formalism to complex
scalar fields, giving particular attention to the effects of a
non-zero chemical potential, which turns out to be non-trivial. In
section II we obtain the structure of the matrix of 2-point functions
in the closed-time-path formalism, which is applied in section III to
find the most general permissible form of the dissipative counterterm.
The corresponding unperturbed propagators are derived in section IV,
where we find that, in the presence of a non-zero chemical potential,
not only occupation numbers but also decay widths and thermal masses
are different for the quasiparticle modes corresponding to particles
and antiparticles. In section V we evaluate the dissipative
counterterm at the lowest non-trivial order and find, with suitable
approximations, that the time-dependent occupation numbers obey a
kinetic equation of the Boltzmann type. Finally, our results are
summarised in section VI.
%%%%%%%%%%%%%%%%%%%%%%%%%%%%%%%%%%%%%%%%%%%%%%%
%  SECTION 2
%%%%%%%%%%%%%%%%%%%%%%%%%%%%%%%%%%%%%%%%%%%%%%%
\section{PROPAGATORS IN THE CLOSED-TIME-PATH FORMALISM}
We consider the theory of a massive, complex, self-interacting scalar
field, $\phi({\bf x},t)$, described by the Lagrangian density
%%%%%%%%%%%%%%%%%%%%%%%%%%%%%%%%%%%%%%%%%%%%%%%
\begin{equation}
{\cal L}(\phi,\phi^{\ast})=\dot{\phi^*}\dot{\phi}-
\nabla\phi^*\nabla\phi-m^{2}(t)\phi^*
\phi- \frac{\lambda}{4}(\phi^*\phi)^2                  \label{eq:Lag}
\end{equation}
%%%%%%%%%%%%%%%%%%%%%%%%%%%%%%%%%%%%%%%%%%%%%%%
where $\dot{\phi}=\partial\phi/\partial t$. The generating
functional for the time ordered Green's functions of the theory may be
written as
%%%%%%%%%%%%%%%%%%%%%%%%%%%%%%%%%%%%%%%%%%%%%%%
\begin{equation}
Z[j]=Tr\left[\rho T{e^{ i\int d^{4}x(j\phi^*+ j^*\phi)}}\right]
                                                    \label{eq:genfunc}
\end{equation}
%%%%%%%%%%%%%%%%%%%%%%%%%%%%%%%%%%%%%%%%%%%%%%%
where $\phi({\bf x},t)$ is the Heisenberg-picture field operator,
$j({\bf x},t)$ is a complex source and $T$ denotes time ordering. For
an initial state of thermal equilibrium at time $t_0$, characterised
by the temperature $1/\beta$ and chemical potential $\mu$, the density
operator is
%%%%%%%%%%%%%%%%%%%%%%%%%%%%%%%%%%%%%%%%%%%%%%%
\begin{equation}
\rho={{e^{-\beta\left[H(t_0)-\mu N\right]}}\over{
Tr\left[e^{-\beta\left[H(t_0)-\mu N\right]}\right]}}\ ,
                                                        \label{eq:rho}
\end{equation}
%%%%%%%%%%%%%%%%%%%%%%%%%%%%%%%%%%%%%%%%%%%%%%%
where $H(t)$ is the time-dependent Hamiltonian and $N$ is the particle
number operator. It is convenient to generalise this generating
functional, using the Schr\"{o}dinger picture field operator
$\phi_s({\bf x})$ to write
%%%%%%%%%%%%%%%%%%%%%%%%%%%%%%%%%%%%%%%%%%%%%%%
\begin{equation}
Z[j_{1},j_{2}] = Tr\left[\rho\bar{T}{e^{i\int d^4x(H_{s}
+j_{2}\phi^*_{s}+j_{2}^*\phi_{s})}} T{e^{-i\int d^4x(H_{s}-
j_{1}\phi^{\ast}_{s}-j_{1}^{\ast}\phi_{s})}}\right]
                                                    \label{eq:schrgen}
\end{equation}
%%%%%%%%%%%%%%%%%%%%%%%%%%%%%%%%%%%%%%%%%%%%%%%
where $\bar{T}$ denotes anti-time ordering and $H_s(t)$ is the
Hamiltonian expressed in terms of $\phi_s$.  This Hamiltonian depends
explicitly on time through the time-dependent mass $m(t)$.  The
original generating functional (\ref{eq:genfunc}) is given by
$Z[j]=Z[j,0]$.

Each of the three factors inside the trace in (\ref{eq:schrgen}) has
a path-integral representation with its own integration variable, say
$\phi_1({\bf x},t)$ for the time-ordered factor, $\phi_2({\bf x},t)$
for the anti-time-ordered factor and $\phi_3({\bf x},\tau)$ for the
density operator, where the real time $t$ runs from the initial time
$t_0$ to some large final time $t_f$, while the imaginary time $\tau$
runs from 0 to $\beta$. As usual, \cite{semenoff,idl:prd}, these three
fields can be envisaged as living on the three segments of a contour
in the complex time plane illustrated in \ref{Fig1}. The effect of a
non-zero chemical potential can formally be taken into account by
using $H(t_0)$ as the generator of evolution in imaginary time and
imposing the boundary conditions
%%%%%%%%%%%%%%%%%%%%%%%%%%%%%%%%%%%%%%%%%%%%%%%
\begin{equation}
\phi_3({\bf x},\beta) = e^{-\beta\mu}\phi_1({\bf x},t_0), \qquad
\phi_3^*({\bf x},\beta) = e^{\beta\mu}\phi^*_1({\bf x},t_0)\ .
                                                   \label{eq:bcsonphi}
\end{equation}
%%%%%%%%%%%%%%%%%%%%%%%%%%%%%%%%%%%%%%%%%%%%%%%
On introducing a third source for $\phi_3$, the generating functional
(\ref{eq:schrgen}) is then given by the path integral
%%%%%%%%%%%%%%%%%%%%%%%%%%%%%%%%%%%%%%%%%%%%%%%
\begin{equation}
Z[j_1,j_2,j_3]= {\cal N}\int[d\phi_1][d\phi_2][d\phi_3]
e^{iS(\phi_1,\phi_2,\phi_3)}                        \label{eq:pathint}
\end{equation}
%%%%%%%%%%%%%%%%%%%%%%%%%%%%%%%%%%%%%%%%%%%%%%%
where ${\cal N}$ is a normalizing constant and
%%%%%%%%%%%%%%%%%%%%%%%%%%%%%%%%%%%%%%%%%%%%%%%
\begin{eqnarray}
S(\phi_1,\phi_2,\phi_3)&=&\int_{t_0}^{t_f}dt \int d^3x\ \left[
{\cal L}(\phi_1) - {\cal L}(\phi_2) + j_1\phi_1^* + j_1^*\phi_1 +
j_2\phi_2^* + j_2^*\phi_2 \right] \nonumber \\
&&\qquad+\int_0^{\beta} d\tau \int d^3x\ \left[i{\cal L}_E(\phi_3) +
j_3\phi_3^* + j_3^*\phi_3\right]\ .                  \label{eq:ctpact}
\end{eqnarray}
%%%%%%%%%%%%%%%%%%%%%%%%%%%%%%%%%%%%%%%%%%%%%%%
Here, ${\cal L}_E$ denotes the euclidean version of (\ref{eq:Lag}).

In order to carry out perturbation theory, we wish to split the
action into an unperturbed part, $S_0$, an interaction part $S_I$ and
a source term $S_j$:
%%%%%%%%%%%%%%%%%%%%%%%%%%%%%%%%%%%%%%%%%%%%%%%
\begin{equation}
S(\phi_1,\phi_2, \phi_3)= S_0(\phi_1,\phi_2, \phi_3)+
S_I(\phi_1,\phi_2, \phi_3)+ S_j(\phi_1,\phi_2, \phi_3)\ .
                                                     \label{eq:action}
\end{equation}
%%%%%%%%%%%%%%%%%%%%%%%%%%%%%%%%%%%%%%%%%%%%%%%
In the standard way, the generating functional (\ref{eq:pathint}) can
then be rewritten as
%%%%%%%%%%%%%%%%%%%%%%%%%%%%%%%%%%%%%%%%%%%%%%%
\begin{eqnarray}
Z[j_1,j_2,j_3]&=&e^{iS_{I}(-i\delta/\delta j_1,-i\delta/\delta j_2,
-i\delta/\delta j_3)}Z_0[j_1,j_2,j_3]\\
Z_0[j_1,j_2,j_3]&=& \exp\left[-\int d^4x d^4y\
j_a^*(x)g_{ab}(x,y)j_b(y)\right]\ .                \label{eq:splitgen}
\end{eqnarray}
%%%%%%%%%%%%%%%%%%%%%%%%%%%%%%%%%%%%%%%%%%%%%%%
The unperturbed propagator $g(x,y)$ satisfies
%%%%%%%%%%%%%%%%%%%%%%%%%%%%%%%%%%%%%%%%%%%%%%%
\begin{equation}
{\cal D}_{k}(x,\overrightarrow{\partial_x}) g(x,y) =
-i\delta(x-y)=g(x,y){\cal D}(y,-\overleftarrow{\partial_y})
                                                         \label{eq:Dg}
\end{equation}
%%%%%%%%%%%%%%%%%%%%%%%%%%%%%%%%%%%%%%%%%%%%%%%
where ${\cal D}(x,\partial)$ is the differential operator
corresponding to $S_0$, i.e. $S_0 = \int d^4x\ \phi_a^*(x){\cal
D}_{ab}(x,\partial) \phi_b(x)$.

The perturbation theory which results from choosing $S_0$ to be simply
the quadratic part $S^{(2)}$ of $S$ was derived some time ago by
Semenoff and Weiss \cite{semenoff} for a theory with somewhat more
general time dependence than (\ref{eq:Lag}). It has the disadvantage
that coefficients in $g(x,y)$ which correspond roughly to particle
occupation numbers are fixed at their initial values, and do not
change so as to reflect the evolving state of the system. As shown by
one of us for the case of a real scalar field \cite{idl:prd,idl:jpa},
this situation can be improved by choosing instead
%%%%%%%%%%%%%%%%%%%%%%%%%%%%%%%%%%%%%%%%%%%%%%%
\begin{equation}
S_0(\phi_1,\phi_2,\phi_3)=S^{(2)}(\phi_1,\phi_2,\phi_3)
+\int d^4x \phi^*_a M_{ab}(x,\partial)\phi_b        \label{eq:modactn}
\end{equation}
%%%%%%%%%%%%%%%%%%%%%%%%%%%%%%%%%%%%%%%%%%%%%%%
where $M$ is a differential operator, to be chosen in such a way that
the unperturbed propagator $g(x,y)$ mimics as nearly as possible the
dissipative behaviour of the full propagator $G(x,y)$.  To make
perturbation theory tractable, we require the term involving $M$ to be
local, so it can involve only the real-time fields $\phi_1$ and
$\phi_2$. Since (as is worth emphasising) we do not wish to change the
overall theory, the interaction term $S_I$ now includes a counterterm
$-\phi^*_a M_{ab}\phi_b$ in addition to the original interaction
$-(\lambda/4)[(\phi_1^*\phi_1)^2 - (\phi_2^*\phi_2)^2]$ and $M$ will
be chosen so that this counterterm subtracts some part of the loop
contributions to the full propagator. In this way, we partially resum
the absorptive parts of these loop contributions, and optimise
$g(x,y)$ as an approximation to $G(x,y)$. In particular, we will find
that $g(x,y)$ now involves quasiparticle occupation numbers which
evolve with time in the expected way.

In order to determine the permissible form of $M$, we first
investigate the structure of the full propagator, whose real-time
components ($a,b = 1, 2$) are given by
%%%%%%%%%%%%%%%%%%%%%%%%%%%%%%%%%%%%%%%%%%%%%%%
\begin{eqnarray}
G_{ab}(x,x')&=&\left.-\frac {\partial}{\partial j^*_a(x)}\frac
{\partial}{\partial j_b(x')}Z[j_1,j_2,j_3]\right\vert_{j=0}
\nonumber \\
&=&\left[ \matrix{ \langle T [\phi(x,t) \phi^*
(x',t')]\rangle_{\mu} &
\langle \phi^* (x',t') \phi (x,t)\rangle_{\mu} \cr
\langle \phi (x,t) \phi^{\ast} (x', t') \rangle_{\mu} &
\langle \bar{T}
[\phi(x,t) \phi^* (x',t')]\rangle_{\mu} \cr} \right]
                                                       \label{eq:2ptf}
\end{eqnarray}
%%%%%%%%%%%%%%%%%%%%%%%%%%%%%%%%%%%%%%%%%%%%%%%
where $\langle\cdots\rangle_{\mu}$ indicates a thermal average in the
ensemble with chemical potential $\mu$. Writing
%%%%%%%%%%%%%%%%%%%%%%%%%%%%%%%%%%%%%%%%%%%%%%%
\begin{equation}
\langle \phi(x)\phi^*(x')\rangle_{\mu} ={\cal H}_{\mu}(x,x')
\Theta(t-t')+{\cal K}_{\mu}(x,x')\Theta(t'-t)           \label{defexp}
\end{equation}
%%%%%%%%%%%%%%%%%%%%%%%%%%%%%%%%%%%%%%%%%%%%%%%
it is simple to show from the Hermiticity of $\rho$ that
${\cal K}_{\mu}(x,x')={\cal H}^*_{\mu}(x',x)$ and from the charge
conjugation (C) symmetry of (\ref{eq:Lag}) that
%%%%%%%%%%%%%%%%%%%%%%%%%%%%%%%%%%%%%%%%%%%%%%%
\begin{equation}
\langle \phi^*(x')\phi(x)\rangle_{\mu} = {\cal H}^*_{-\mu}(x,x')
\Theta(t-t') + {\cal H}_{-\mu}(x',x)\Theta(t'-t) .
\end{equation}
%%%%%%%%%%%%%%%%%%%%%%%%%%%%%%%%%%%%%%%%%%%%%%%
For any complex function of $\mu$, we define the $\sharp$-conjugate
$f^{\sharp}(\mu) = f^*(-\mu)$, which is related to charge conjugation.
With this notation, the full propagator can be expressed as
%%%%%%%%%%%%%%%%%%%%%%%%%%%%%%%%%%%%%%%%%%%%%%%
\begin{eqnarray}
G(x,x')&=& \left[\matrix{
{\cal H}(x,x') & {\cal H}^{\sharp}(x,x') \cr
{\cal H}(x,x') & {\cal H}^{\sharp}(x,x') \cr}\right]
\Theta (t-t')  \nonumber \\
&&\qquad + \left[\matrix{{\cal H}^{* \sharp}(x',x) &
{\cal H}^{* \sharp}(x',x) \cr
{\cal H}^{*}(x',x) & {\cal H}^{*}(x',x) \cr} \right] \Theta (t'-t)
                                                    \label{eq:fulprop}
\end{eqnarray}
%%%%%%%%%%%%%%%%%%%%%%%%%%%%%%%%%%%%%%%%%%%%%%%
where the suffix $\mu$ has been suppressed.  This structure should be
maintained if the scalar field is embedded in a more general C- or
CP-invariant theory.
%%%%%%%%%%%%%%%%%%%%%%%%%%%%%%%%%%%%%%%%%%%%%%%%%%%
% SECTION 3
%%%%%%%%%%%%%%%%%%%%%%%%%%%%%%%%%%%%%%%%%%%%%%%%%%%%
\section{THE DISSIPATIVE COUNTERTERM}
We now want to construct a differential operator ${\cal D}$ and the
corresponding counterterm matrix $M$ in such a way that equation
(\ref{eq:Dg}) admits a solution for $g(x,y)$ of the form
(\ref{eq:fulprop}). Since we are dealing with a spatially homogeneous
theory, we can take a spatial Fourier transform, obtaining a
propagator $g_k(t,t')$, and operator ${\cal D}_k(t,\partial_t)$ which
obey (\ref{eq:Dg}) in the form
%%%%%%%%%%%%%%%%%%%%%%%%%%%%%%%%%%%%%%%%%%%%%%%
\begin{equation}
{\cal D}_k(t,\overrightarrow{\partial_t})g_k(t,t')=-i\delta(t-t')
=g_k(t,t'){\cal D}_k(t',-\overleftarrow{\partial_{t'}}) \ .
                                                        \label{eq:Dgk}
\end{equation}
%%%%%%%%%%%%%%%%%%%%%%%%%%%%%%%%%%%%%%%%%%%%%%%
In the following, we will usually suppress the suffix $k$. As in
\cite{idl:prd}, we first construct the operator ${\cal D}(\partial_t)$
with constant coefficients appropriate to a temporally homogeneous
system, and then allow these coefficients to depend on time. In the
temporally homogeneous case, $g(t,t')$ can be expressed in the form of
(\ref{eq:fulprop}) in terms of a function $h(t-t')$ (which also
depends on $k$ and $\mu$) for which we write
%%%%%%%%%%%%%%%%%%%%%%%%%%%%%%%%%%%%%%%%%%%%%%%
\begin{eqnarray}
h(t-t')&=&u(t-t')+iv(t-t') \\
h^{* \sharp}(t-t')&=&w(t-t')+iz(t-t')                 \label{eq:upliv}
\end{eqnarray}
%%%%%%%%%%%%%%%%%%%%%%%%%%%%%%%%%%%%%%%%%%%%%%%
where $u$, $v$, $w$, and $z$ are real functions. (Note that $f^{*
\sharp}(\mu) = f(-\mu)$ for any function $f(\mu)$ ). For the temporal
Fourier transform
%%%%%%%%%%%%%%%%%%%%%%%%%%%%%%%%%%%%%%%%%%%%%%%
\begin{equation}
\hat{g}(\omega)= \int^{\infty}_{-\infty}dt e^{i\omega t}g(t)
                                                         \label{eq:ft}
\end{equation}
%%%%%%%%%%%%%%%%%%%%%%%%%%%%%%%%%%%%%%%%%%%%%%%
we obtain
\begin{eqnarray}
\hat{g}_{11}(\omega)&=&\hat{g}_{22}^*(\omega) \nonumber \\
&=& \left[A(\omega)-b(\omega)+C(\omega)+d(\omega)\right]\nonumber \\
&&\quad+i\left[a(\omega)+B(\omega)- c(\omega)+D(\omega)\right] \\
\hat{g}_{12}(\omega)&=&2\left[C(\omega)+d(\omega)\right] \nonumber \\
\hat{g}_{21}(\omega)&=&2\left[A(\omega)-b(\omega)\right] \nonumber
\end{eqnarray}
%%%%%%%%%%%%%%%%%%%%%%%%%%%%%%%%%%%%%%%%%%%%%%%
where
%%%%%%%%%%%%%%%%%%%%%%%%%%%%%%%%%%%%%%%%%%%%%%%
\begin{eqnarray*}
A(\omega)+ia(\omega)&=&\int_{0}^{\infty} dt u(t) \left[
\cos \omega t +i\sin \omega t\right] \\
B(\omega)+ib(\omega)&=&\int_{0}^{\infty} dt v(t) \left[
\cos \omega t +i\sin \omega t\right] \\
C(\omega)+ic(\omega)&=&\int_{0}^{\infty} dt w(t) \left[
\cos \omega t +i\sin \omega t\right] \\
D(\omega)+id(\omega)&=&\int_{0}^{\infty} dt z(t) \left[
\cos \omega t +i\sin \omega t\right] \ .
\end{eqnarray*}
%%%%%%%%%%%%%%%%%%%%%%%%%%%%%%%%%%%%%%%%%%%%%%%
Note that the change of variables $(\omega,\mu)\to(-\omega,-\mu)$
leads to the interchange $(A(\omega),a(\omega),B(\omega),b(\omega))
\leftrightarrow (C(\omega),-c(\omega),D(\omega),-d(\omega))$, and
consequently
%%%%%%%%%%%%%%%%%%%%%%%%%%%%%%%%%%%%%%%%%%%%%%%
\begin{equation}
\hat{g}_{ab}(\omega)=\hat{g}_{ba}^{* \sharp}(-\omega) \label{eq:musym}
\end{equation}
%%%%%%%%%%%%%%%%%%%%%%%%%%%%%%%%%%%%%%%%%%%%%%%
which also follows directly from (\ref{eq:fulprop}).  Upon Fourier
transformation, (\ref{eq:Dgk}) becomes an algebraic equation whose
solution for ${\cal D}(-i\omega)$ is
%%%%%%%%%%%%%%%%%%%%%%%%%%%%%%%%%%%%%%%%%%%%%%%
\begin{equation}
{\cal D}(-i\omega)={{-i}\over{\det\vert \hat{g}\vert}}
\left[\matrix{\hat{g}_{22}(\omega)&-\hat{g}_{12}(\omega)\cr
-\hat{g}_{21}(\omega)&\hat{g}_{11}(\omega)\cr}\right] \ .
                                                        \label{Dcmpts}
\end{equation}
%%%%%%%%%%%%%%%%%%%%%%%%%%%%%%%%%%%%%%%%%%%%%%%
This solution shows how ${\cal D}$ can be constructed from the
(unknown) functions $A(\omega)\cdots D(\omega)$ which are real, even
functions of $\omega$ and $a(\omega)\cdots d(\omega)$ which are real
and odd. We will choose these eight functions in such a way that
${\cal D}$ has certain essential properties. First, we require ${\cal
D}(\partial_t)$ to be a second-order differential operator. The
$\delta(t-t')$ in (\ref{eq:Dgk}) arises from the derivatives of
$\Theta(t-t')$ and $\Theta(t'-t)$ and, in order that these appear only
in the diagonal elements, second derivatives may appear only in the
diagonal elements of ${\cal D}(\partial_t)$. To maintain the
normalization of $\phi$ as in (\ref{eq:Lag}), we fix the coefficients
of $\partial_t^2$, or of $-\omega^2$, to be $\pm 1$. Expanding
(\ref{Dcmpts}) in powers of $\omega$, we find that the most general
form of ${\cal D}$ satisfying these requirements can be expressed in
terms of six real coefficients $\alpha$, $\beta$, $\gamma$,
$\bar{\alpha}$, $\bar{\gamma}$, $\tilde{\gamma}$ as
%%%%%%%%%%%%%%%%%%%%%%%%%%%%%%%%%%%%%%%%%%%%%%%
\begin{equation}
{\cal D}(-i\omega) = \left[\matrix{
-\omega^2-i(\bar{\gamma}-i\tilde{\gamma})\omega+ \beta -i\alpha &
-i(\gamma-\bar{\gamma})\omega +i(\alpha+\overline{\alpha}) \cr
i(\gamma+\bar{\gamma})\omega +i(\alpha-\bar{\alpha}) &
\omega^2 -i(\bar{\gamma}+i\tilde{\gamma})\omega - \beta-i\alpha}
\right] \ .                                           \label{Dofomega}
\end{equation}
%%%%%%%%%%%%%%%%%%%%%%%%%%%%%%%%%%%%%%%%%%%%%%%
Since ${\cal D}(-i\omega)$ is equal to $-ig^{-1}(\omega)$, it has the
same symmetry (\ref{eq:musym}) as $g(\omega)$, and we may deduce that
$\alpha$, $\beta$ and $\gamma$ are even functions of $\mu$, while
$\bar{\alpha}$, $\bar{\gamma}$ and $\tilde{\gamma}$ are odd.  Since
these quantities are also real, we have
%%%%%%%%%%%%%%%%%%%%%%%%%%%%%%%%%%%%%%%%%%%%%%%
\begin{eqnarray}
\alpha^{\sharp} = \alpha,\quad \beta^{\sharp}&=&\beta, \quad
\gamma^{\sharp} = \gamma \nonumber\\
\bar{\alpha}^{\sharp} = - \bar{\alpha},\quad \bar{\gamma}^{\sharp}&=&-
\bar{\gamma}, \quad \tilde{\gamma}^{\sharp} = - \tilde{\gamma} \ .
                                                   \label{eq:alphasym}
\end{eqnarray}
%%%%%%%%%%%%%%%%%%%%%%%%%%%%%%%%%%%%%%%%%%%%%%%
In particular, when $\mu$ vanishes, we have $\bar{\alpha}=\bar{\gamma}
=\tilde{\gamma}=0$ and ${\cal D}$ reduces to the form found in
\cite{idl:prd} for a real scalar field.

For the non-equilibrium theory, with time-dependent bare mass, ${\cal
D}(t,\partial_t)$ is given by
%%%%%%%%%%%%%%%%%%%%%%%%%%%%%%%%%%%%%%%%%%%%%%%
\begin{equation}
\begin{array}{l}
{\cal D}_{11}=
\partial_{t}^{2}+(\bar{\gamma}-i\tilde{\gamma})\partial_t+ \beta
-i\alpha +(\dot{\bar{\gamma}}-i\dot{\tilde{\gamma}})/2
\\
{\cal D}_{12}=
(\gamma-\bar{\gamma})\partial_t +i(\alpha+\bar{\alpha})+(\dot{\gamma}
-\dot{\bar{\gamma}})/2
\\
{\cal D}_{21}=
-(\gamma+\bar{\gamma})\partial_t+i(\alpha-\bar{\alpha}) -(\dot{\gamma}
+\dot{\bar{\gamma}})/2
\\
{\cal D}_{22}=
-\partial_t^{2}+(\bar{\gamma}+i\tilde{\gamma})\partial_t - \beta
-i\alpha +(\dot{\bar{\gamma}}+i\dot{\tilde{\gamma}})/2
\end{array}                                            \label{eq:Doft}
\end{equation}
%%%%%%%%%%%%%%%%%%%%%%%%%%%%%%%%%%%%%%%%%%%%%%%
where the coefficients are now time dependent, and the derivatives
$\dot{\gamma}$, etc ensure that ${\cal D}(t,\partial_t)$ is a
symmetrical operator, as required in (\ref{eq:Dg}). That is,
$\gamma\partial_t + \dot{\gamma}/2 = \gamma^{1/2}\partial_t
\gamma^{1/2}$, etc.  We emphasise again that the six coefficients in
(\ref{eq:Doft}), while undetermined at this stage, do not represent
arbitrary modifications of the theory, but rather an optimal choice of
the lowest-order theory which is to serve as the basis for
perturbation theory. As indicated above, these coefficients will be
found self-consistently from a suitable renormalization prescription,
and we shall find that they correspond to six properties of a gas of
quasiparticles, namely the thermal masses, occupation numbers and
relaxation rates of particles and antiparticles.
%%%%%%%%%%%%%%%%%%%%%%%%%%%%%%%%%%%%%%%%%%%%%%%%%%%%%%%%%%%
% SECTION 4
%%%%%%%%%%%%%%%%%%%%%%%%%%%%%%%%%%%%%%%%%%%%%%%%%%%%%%%%%%%
\section{THE DISSIPATIVE PROPAGATOR}
The basic equation (\ref{eq:Dgk}) for the unperturbed propagator
matrix $g(t,t')$ is satisfied if the function $h(t,t')$ obeys the
conditions
%%%%%%%%%%%%%%%%%%%%%%%%%%%%%%%%%%%%%%%%%%%%%%%
\begin{eqnarray}
\partial_t\left.\left[h(t,t') - h^{* \sharp}(t',t)\right]\right\vert_
{t'=t}&=&-i \\
\partial_t\left.\left[h(t,t')-h^*(t',t)\right]\right\vert_{t'=t}
&=&0 \\
\left[{\cal D}_{11}(t,\partial_t) + {\cal D}_{12}(t,\partial_t)\right]
h(t,t')&=&0 \\
{\cal D}_{11}(t,\partial_t)h^{* \sharp}(t',t) + {\cal D}_{12}(t,
\partial_t)h^*(t',t)&=&0                           \label{eq:eqnsforh}
\end{eqnarray}
%%%%%%%%%%%%%%%%%%%%%%%%%%%%%%%%%%%%%%%%%%%%%%%
together with their $\sharp$-conjugates and the continuity conditions
%%%%%%%%%%%%%%%%%%%%%%%%%%%%%%%%%%%%%%%%%%%%%%%
\begin{equation}
h(t,t)=h^*(t,t)=h^{\sharp}(t,t) \ .
\end{equation}
%%%%%%%%%%%%%%%%%%%%%%%%%%%%%%%%%%%%%%%%%%%%%%%
We find that the solution of these
equations can be written in the form
%%%%%%%%%%%%%%%%%%%%%%%%%%%%%%%%%%%%%%%%%%%%%%%
\begin{eqnarray}
h(t,t')&=&{{e^{-\frac{1}{2}\int_{t'}^t\left[\gamma(\bar{t})
-i\tilde{\gamma}(\bar{t})\right]d\bar{t}}}\over{4\sqrt{\Omega(t)
\Omega(t')}}}\left[\left(1+N(t')\right)e^{-i\int_{t'}^t\Omega(\bar{t})
d\bar{t}}\right. \nonumber \\
&&\qquad\qquad\qquad\qquad\qquad +\left.\left(-1+N^{\sharp}(t')\right)
e^{i\int_{t'}^t\Omega(\bar{t}) d\bar{t}}\right]\ ,    \label{eq:hsoln}
\end{eqnarray}
%%%%%%%%%%%%%%%%%%%%%%%%%%%%%%%%%%%%%%%%%%%%%%%
in terms of the auxiliary functions $\Omega(t)$ and $N(t)$ which are
solutions of
%%%%%%%%%%%%%%%%%%%%%%%%%%%%%%%%%%%%%%%%%%%%%%%
\begin{equation}
\frac{1}{2}\frac{\ddot{\Omega}}{\Omega} - \frac{3}{4}
\frac{\dot{\Omega}^2}{\Omega^2}+\Omega^2 = \beta -\frac{1}{4}(\gamma -
i\tilde{\gamma})^2 + i\bar{\alpha}                 \label{eq:Omegaeqn}
\end{equation}
\begin{equation}
\left[\partial_t +2i\Omega - \frac{\dot{\Omega}}{\Omega} +\gamma
\right]\left[(\partial_t + \gamma)N -\bar{\gamma}\right]
= 2i\alpha +2i\bar{\alpha}N+ i\tilde{\gamma}(\gamma N - \bar{\gamma})
                                                       \label{eq:Neqn}
\end{equation}
%%%%%%%%%%%%%%%%%%%%%%%%%%%%%%%%%%%%%%%%%%%%%%%
with the subsidiary conditions
%%%%%%%%%%%%%%%%%%%%%%%%%%%%%%%%%%%%%%%%%%%%%%%
\begin{equation}
\left({{N+N^{\sharp}}\over\Omega}\right)^*=
\left({{N+N^{\sharp}}\over\Omega}\right)                 \label{eq:Na}
\end{equation}
\begin{eqnarray}
\frac{d}{dt}\left(\frac{N+N^{\sharp}}{\Omega}\right)
&=&-\frac{1}{2}
\left(\frac{\dot{\Omega}}{\Omega} + \frac{\dot{\Omega}^*}{\Omega^*}
+2\gamma\right)\left(\frac{N+N^{\sharp}}{\Omega}\right)\nonumber\\
&&\qquad\qquad
-i\left[(N-N^{\sharp})-(N-N^{\sharp})^*\right] \ .       \label{eq:Nb}
\end{eqnarray}
%%%%%%%%%%%%%%%%%%%%%%%%%%%%%%%%%%%%%%%%%%%%%%%
It is straightforward, though tedious, to show that these conditions
are preserved by (\ref{eq:Neqn}) if they hold at the initial time.

With a non-zero chemical potential, $\tilde{\gamma}$ and
$\bar{\alpha}$ are in general nonzero, so the frequency $\Omega(t)$
which satisfies (\ref{eq:Omegaeqn}) is complex, though it has the
property $\Omega^{\sharp} = \Omega$.

To obtain the initial conditions that apply to (\ref{eq:Omegaeqn}) and
(\ref{eq:Neqn}), we require the full $3\times 3$ matrix of propagators
which satisfies an equation of the form (\ref{eq:Dgk}) with
%%%%%%%%%%%%%%%%%%%%%%%%%%%%%%%%%%%%%%%%%%%%%%%
\begin{equation}
{\cal D}_{33}(\partial_{\tau}) = i(\partial_{\tau}^2 -
\omega^2)                                               \label{eq:D33}
\end{equation}
%%%%%%%%%%%%%%%%%%%%%%%%%%%%%%%%%%%%%%%%%%%%%%%
and ${\cal D}_{13}={\cal D}_{23}={\cal D}_{31}={\cal D}_{32}=0$, where
$\omega^2 = k^2 + m^2(t_0)$. The boundary conditions which apply to
these nine propagators are set out in the Appendix, where results for
those involving imaginary times are also given.

For the case of a time-independent mass $m^2$, we expect a steady
state solution, with $\dot{\Omega}=\dot{N} = 0$.  In this case, we
find that the boundary conditions can be satisfied only if three
relations hold between $\alpha$, $\beta$, $\cdots$ . The first of
these is
%%%%%%%%%%%%%%%%%%%%%%%%%%%%%%%%%%%%%%%%%%%%%%%
\begin{equation}
\tilde{\gamma}^2 = 4(\beta - \omega^2)\ .             \label{eq:gteqn}
\end{equation}
%%%%%%%%%%%%%%%%%%%%%%%%%%%%%%%%%%%%%%%%%%%%%%%
It seems natural to require that the renormalized masses (or, more
generally, the $k$-dependent frequencies) in ${\cal D}_{11}$ and
${\cal D}_{33}$ should be equal, although this is not obligatory.
This means that $\beta = \omega^2 = m^2 + k^2$, and hence that
$\tilde{\gamma} = 0$.  We will indeed assume that $\tilde{\gamma} =
0$, since this introduces considerable simplifications, but the
consequences of this assumption will need some discussion at a later
stage.  In this case, the two other relations are
%%%%%%%%%%%%%%%%%%%%%%%%%%%%%%%%%%%%%%%%%%%%%%%
\begin{eqnarray}
\frac{1}{2}\left[\alpha + \bar{\alpha} -(\gamma - \bar{\gamma})\omega
\right] - (\gamma\omega - \bar{\alpha})n^- = 0     \label{eq:eqrel1}\\
\frac{1}{2}\left[\alpha - \bar{\alpha} -(\gamma + \bar{\gamma})\omega
\right] - (\gamma\omega + \bar{\alpha})n^+ = 0      \label{eq:eqrel2}
\end{eqnarray}
%%%%%%%%%%%%%%%%%%%%%%%%%%%%%%%%%%%%%%%%%%%%%%%
where
%%%%%%%%%%%%%%%%%%%%%%%%%%%%%%%%%%%%%%%%%%%%%%%
\begin{equation}
n^{\pm}=\frac{1}{e^{\beta(\omega \pm \mu)}-1}          \label{eq:beno}
\end{equation}
%%%%%%%%%%%%%%%%%%%%%%%%%%%%%%%%%%%%%%%%%%%%%%%
are the usual occupation numbers for particles and antiparticles in a
free Bose-Einstein gas.  The coefficient $N$ in the real-time
propagators is then given by
%%%%%%%%%%%%%%%%%%%%%%%%%%%%%%%%%%%%%%%%%%%%%%%
\begin{equation}
N=\left(\frac{\Omega + i\gamma/2}{\omega}\right)\left(1+n^++n^-\right)
+\left(n^--n^+\right)                                \label{eq:Nequil}
\end{equation}
%%%%%%%%%%%%%%%%%%%%%%%%%%%%%%%%%%%%%%%%%%%%%%%%
and the frequency $\Omega$ by
%%%%%%%%%%%%%%%%%%%%%%%%%%%%%%%%%%%%%%%%%%%%%%%
\begin{equation}
\Omega^2 = \omega^2 - \gamma^2/4 + i\bar{\alpha}\ . \label{eq:Omequil}
\end{equation}
%%%%%%%%%%%%%%%%%%%%%%%%%%%%%%%%%%%%%%%%%%%%%%%
Note that when the dissipative counterterm is neglected, we have
$\Omega=\omega$ and $\gamma = 0$, so that $N=1+2n^-$.  We then find,
as expected, that the coefficient of the positive frequency term in
the propagator (\ref{eq:hsoln}) is $(1+N)=2(1+n^-)$ while that of the
negative frequency term is $(-1+N^{\sharp})=2n^+$.  Thus, with the
dissipative terms present, the real-time propagators describe a gas of
quasiparticles, in which quasiparticle modes contain a small admixture
of bare antiparticles and quasi-antiparticle modes contain a small
admixture of bare particles.

For this gas of quasiparticles, the positive- and negative-frequency
mode functions are $\exp[-i\Omega_+(t-t')]$ and
$\exp[i\Omega_-(t-t')]$ respectively, where
%%%%%%%%%%%%%%%%%%%%%%%%%%%%%%%%%%%%%%%%%%%%%%%
\begin{equation}
\Omega_{\pm}=\left[{\rm Re}\Omega \mp\tilde{\gamma}/2\right]\mp
i\left[\gamma/2 \mp {\rm Im}\Omega\right]\ .        \label{eq:Omegapm}
\end{equation}
%%%%%%%%%%%%%%%%%%%%%%%%%%%%%%%%%%%%%%%%%%%%%%%
We see that the decay rates for particle and antiparticle modes will
be different if ${\rm Im}\Omega\ne 0$, which will in general be true
if either $\tilde{\gamma}$ or $\bar{\alpha}$ is nonzero. On the other
hand, the thermal masses of quasiparticles and quasi-antiparticles
will be different if and only if $\tilde{\gamma}$ is nonzero.  Which
of these conditions actually applies will be discussed in the next
section.

For the non-equilibrium theory, any solution of (\ref{eq:Omegaeqn})
may in principle be used for $\Omega(t)$. Clearly, however, it is
desirable that our lowest-order theory should approximately retain the
characteristics of the steady-state solution in the case where
$m^2(t)$ is slowly varying near $t_0$. We will therefore choose a
renormalization prescription for which $\beta(t_0) = \omega^2$, and
the solution of (\ref{eq:Omegaeqn}) which has $\dot{\Omega}(t_0)=0$
and $\Omega(t_0)$ given by (\ref{eq:Omequil}). Retaining the
assumption that $\tilde{\gamma}(t_0)=0$, (\ref{eq:Omegaeqn}) shows
that $\ddot{\Omega}(t_0) = 0$ also. We would now like to interpret
$N(t)$ in terms of time-dependent occupation numbers $n^{\pm}(t)$. To
this end, we define
%%%%%%%%%%%%%%%%%%%%%%%%%%%%%%%%%%%%%%%%%%%%%%%
\begin{eqnarray}
n^{\pm}(t)&=&\frac{1}{4\Omega(t)}\left\{[\surd\beta(t) \pm
i\gamma(t)/2] [N(t)+N^{\sharp}(t)] \right.\nonumber \\
&&\qquad\qquad\qquad \left. \mp \Omega(t)[N(t)-N^{\sharp}(t)]
-2\Omega(t)\right\}                                  \label{eq:npmoft}
\end{eqnarray}
%%%%%%%%%%%%%%%%%%%%%%%%%%%%%%%%%%%%%%%%%%%%%%%
so that
%%%%%%%%%%%%%%%%%%%%%%%%%%%%%%%%%%%%%%%%%%%%%%%
\begin{equation}
N(t) = \left(\frac{\Omega(t)+i\gamma(t)/2}{\surd\beta(t)}\right)\left[
1+n^+(t)+n^-(t)\right] + \left[n^-(t)-n^+(t)\right]\ .
                                                       \label{eq:Noft}
\end{equation}
%%%%%%%%%%%%%%%%%%%%%%%%%%%%%%%%%%%%%%%%%%%%%%%
It is easy to see that $n^+(t)$ and $n^-(t)$ are real and that $n^{+
\sharp}(t) = n^-(t)$ as required for the occupation numbers.  If
$n^{\pm}(t_0)$ are taken to be the equilibrium values (\ref{eq:beno}),
then the desired initial condition $\dot{N}(t_0) = 0$ ensures that
(\ref{eq:Nb}) is satisfied.

As in \cite{idl:prd}, we can now show that with reasonable
approximations, the evolution equation (\ref{eq:Neqn}) and its
$\sharp$-conjugate reduce to a pair of Boltzmann-like kinetic
equations for $n^{\pm}(t)$. We first rewrite (\ref{eq:Neqn}) with
$\tilde{\gamma}=0$ as
%%%%%%%%%%%%%%%%%%%%%%%%%%%%%%%%%%%%%%%%%%%%%%%
\begin{eqnarray}
&\Omega\left[\partial_t + 2i(\Omega-i\gamma/2)\right]P = 2i(\alpha
+\bar{\alpha})N&                                    \label{eq:Neqna}\\
&P=\Omega^{-1}\left[(\partial_t + \gamma)N-\bar{\gamma}\right]&
\end{eqnarray}
%%%%%%%%%%%%%%%%%%%%%%%%%%%%%%%%%%%%%%%%%%%%%%%
and assume that $\partial_tP<<(2\Omega-i\gamma)P$, which will be
valid if the characteristic relaxation time $1/\gamma$ is much greater
than $1/\Omega$.  It will turn out, as in \cite{idl:prd}, that
$\gamma$ is smaller than $\Omega$ by a factor of $\lambda^2$.  Indeed,
$\gamma$, $\bar{\gamma}$, $\alpha$ and $\bar{\alpha}$ are all of order
$\lambda^2$.  Consequently, to order $\lambda^2$, we can replace $N$
by its lowest-order value $N\approx (1+2n^-)$ and neglect $i\gamma/2$
in comparison with $\Omega$. With these approximations,
(\ref{eq:Neqna}) and its $\sharp$-conjugate become
%%%%%%%%%%%%%%%%%%%%%%%%%%%%%%%%%%%%%%%%%%%%%%%
\begin{eqnarray}
\frac{dn^-}{dt} = \frac{1}{2\Omega}\left[\alpha+\bar{\alpha} -(\gamma
- \bar{\gamma})\Omega\right] -\frac{1}{\Omega}(\gamma\Omega
-\bar{\alpha})n^-                                   \label{eq:dn-dt}\\
\frac{dn^+}{dt} = \frac{1}{2\Omega}\left[\alpha-\bar{\alpha} -(\gamma
+ \bar{\gamma})\Omega\right] -\frac{1}{\Omega}(\gamma\Omega
+\bar{\alpha})n^+ \ .                                 \label{eq:dn+dt}
\end{eqnarray}
%%%%%%%%%%%%%%%%%%%%%%%%%%%%%%%%%%%%%%%%%%%%%%%
Evidently, the relations (\ref{eq:eqrel1}) and (\ref{eq:eqrel2}) which
apply to the steady state solution are just the conditions for the
occupation numbers to be constant. In the next section, we evaluate
$\alpha$, $\bar{\alpha}$, $\gamma$ and $\bar{\gamma}$ explicitly, and
find that the right hand sides of (\ref{eq:dn-dt}) and
(\ref{eq:dn+dt}) have approximately the form of the scattering
integrals which appear in the Boltzmann equation.
%%%%%%%%%%%%%%%%%%%%%%%%%%%%%%%%%%%%%%%%%%%%%%%%
% SECTION 5
%%%%%%%%%%%%%%%%%%%%%%%%%%%%%%%%%%%%%%%%%%%%%%%
\section{EVALUATION OF THE DISSIPATIVE COUNTERTERM}
As explained above, our strategy is to choose the counterterm $M$ so
as to optimise the unperturbed propagator $g(x,y)$ as an approximation
to the full propagator $G(x,y)$.  The relation between these may be
expressed by the Dyson-Schwinger equation
%%%%%%%%%%%%%%%%%%%%%%%%%%%%%%%%%%%%%%%%%%%%%%%
\begin{equation}
G_{ab}(x,y)=g_{ab}(x,y)+i\int\int d^{4}z d^{4}z'\ g_{ac}(x,z)
\Sigma_{cd}(z,z') G_{db}(z',y)                       \label{eq:dysons}
\end{equation}
%%%%%%%%%%%%%%%%%%%%%%%%%%%%%%%%%%%%%%%%%%%%%%%
in terms of the self-energy $\Sigma(x,y)$. As explained in detail in
\cite{idl:prd}, we express $\Sigma$ in terms of the average $\bar{t}$
and difference $\Delta t$ of its time arguments, and take the Fourier
transform with respect to $({\bf x} - {\bf y})$ and $\Delta t$. The
result can be written as
%%%%%%%%%%%%%%%%%%%%%%%%%%%%%%%%%%%%%%%%%%%%%%%
\begin{equation}
\Sigma(\bar{t},\omega) = M(\bar{t},-i\omega) +
\tilde{\Sigma}^{(1)}(\bar{t},\omega^2)\omega +
\tilde{\Sigma}^{(2)}(\bar{t}, \omega^2)          \label{eq:splitsigma}
\end{equation}
%%%%%%%%%%%%%%%%%%%%%%%%%%%%%%%%%%%%%%%%%%%%%%%
where $\tilde{\Sigma}$ denotes the contribution from loop diagrams,
which has been split into parts even and odd in the transform variable
$\omega$.  In the counterterm contribution, the $\bar{t}$ dependence
is that of $\alpha(\bar{t})\cdots$ while $-i\omega$ replaces
$\partial_t$. While $M(\bar{t},-i\omega)$ is a second-order polynomial
in $\omega$, the same is not true of $\tilde{\Sigma}(\bar{t},\omega)$.
As in standard renormalization theory, therefore, we choose the
counterterm to cancel $\tilde{\Sigma}$ at some reference frequency,
which can conveniently be chosen as $\bar{\Omega}=\Omega(\bar{t})$.
This gives
%%%%%%%%%%%%%%%%%%%%%%%%%%%%%%%%%%%%%%%%%%%%%%%
\begin{eqnarray}
{[}\gamma(t)-\bar{\gamma}(t)]&=&i\Sigma_{12}^{(1)}(t,\bar{\Omega}^{2})
                                               \label{eq:sigma12(1)}\\
{[}\alpha(t)+\bar{\alpha}(t)]&=&i\Sigma_{12}^{(2)}(t,\bar{\Omega}^{2})
                                               \label{eq:sigma12(2)}\\
{[}\bar{\gamma}(t)- i\tilde{\gamma}(t)]&=&i\Sigma_{11}^{(1)}
(t,\bar{\Omega}^2)
                                                 \label{eq:sigma11(1)}
\end{eqnarray}
%%%%%%%%%%%%%%%%%%%%%%%%%%%%%%%%%%%%%%%%%%%%%%%
from which $\alpha$, $\bar{\alpha}$, $\gamma$, $\bar{\gamma}$ and
$\tilde{\gamma}$ can be deduced by isolating parts which are even and
odd under $\sharp$-conjugation.  The corresponding equation for
$\beta$, namely
%%%%%%%%%%%%%%%%%%%%%%%%%%%%%%%%%%%%%%%%%%%%%%%
\begin{equation}
k^{2}+m^{2}(t)-\beta(t) = {\rm Re}\Sigma_{11}^{(2)}(t,\bar{\Omega}^2)
\end{equation}
%%%%%%%%%%%%%%%%%%%%%%%%%%%%%%%%%%%%%%%%%%%%%%%
can be adjusted to meet the requirement $\beta(t_0)=k^2 + m_R^2$,
where $m_R$ is an appropriate renormalized mass, as discussed in the
last section, but the details are not important for our present
purposes.

The lowest-order contribution to $\Sigma_{12}$ is from the graph shown
in \ref{Fig2}. An approximate method for evaluating it is described in
\cite{idl:prd}.  Since the whole diagram is proportional to
$\lambda^2$, we take the lowest-order approximation to its internal
propagators, using
%%%%%%%%%%%%%%%%%%%%%%%%%%%%%%%%%%%%%%%%%%%%%%%
\begin{equation}
h(t,t')\approx
\frac{1}{2\Omega(\bar{t})}\left[\left(1+n^-(\bar{t})\right)
e^{-i\Omega(\bar{t})\Delta t} + n^+(\bar{t})e^{i\Omega(\bar{t}) \Delta
t} \right]\ ,                                       \label{eq:happrox}
\end{equation}
%%%%%%%%%%%%%%%%%%%%%%%%%%%%%%%%%%%%%%%%%%%%%%%
where $n^{\pm}(\bar{t})$ and $\Omega(\bar{t})$ are assumed to vary
sufficiently slowly that they can be treated as effectively constant.
After calculating $\alpha_k(t)$, $\bar{\alpha}_k(t)$, $\gamma_k(t)$
and $\bar{\gamma}(t)$ from (\ref{eq:sigma12(1)}) -
(\ref{eq:sigma11(1)}) and substituting the results in (\ref{eq:dn-dt})
and (\ref{eq:dn+dt}), we obtain the Boltzmann equation
%%%%%%%%%%%%%%%%%%%%%%%%%%%%%%%%%%%%%%%%%%%%%%%
\begin{eqnarray}
\frac{d}{dt}n_{k}^{-}&\approx&\frac{\lambda^{2}}{32(2\pi)^{5}}\int
d^{3}k_{1} d^{3}k_{2}
d^{3}k_{3}\ \delta({\bf k}_{1}+{\bf k}_{2}-{\bf k}_{3}-{\bf k})
\nonumber \\
&&\quad
\times \delta(\Omega_{1}+\Omega_{2}-\Omega_{3}-\Omega_{k})
\left[\Omega_{1} \Omega_{2}\Omega_{3}\Omega_{k}\right]^{-1}
\nonumber \\
&&\quad
\times \left[n_{1}^{+}n_{2}^{-}(1+n_{3}^{+})(1+n_{k}^{-})
-(1+n_{1}^{+})(1+n_{2}^{-})n_{3}^{+}n_{k}^{-}\right. \nonumber \\
&&\qquad
+n_{1}^{-}n_{2}^{+}(1+n_{3}^{+})(1+n_{k}^{-})
-(1+n_{1}^{-})(1+n_{2}^{+})n_{3}^{+}n_{k}^{-} \nonumber \\
&&\qquad
\left. +n_{1}^{-}n_{2}^{-}(1+n_{3}^{-})(1+n_{k}^{-})
-(1+n_{1}^{-})(1+n_{2}^{-})n_{3}^{-}n_{k}^{-}\right] \ ,
                                                      \label{eq:boltz}
\end{eqnarray}
%%%%%%%%%%%%%%%%%%%%%%%%%%%%%%%%%%%%%%%%%%%%%%%
where $\Omega_1$ denotes $\Omega_{k_1}(t)$, etc. At the order of
approximation we are using, $\Omega_k(t)$ is real. The scattering
integral on the right correctly describes the rate of production minus
the rate of absorption of particles of momentum $k$ due to all 2-body
processes allowed by charge conservation. It vanishes when the
occupation numbers have the Bose-Einstein form
%%%%%%%%%%%%%%%%%%%%%%%%%%%%%%%%%%%%%%%%%%%%%%%
\begin{equation}
n^{\pm}=\frac{1}{e^{\beta(\Omega \pm \bar{\mu})}-1}\ ,
\end{equation}
%%%%%%%%%%%%%%%%%%%%%%%%%%%%%%%%%%%%%%%%%%%%%%%
where the chemical potential $\bar{\mu}$ need not be the same as the
original $\mu$.  Of course, the rate of change of $n^+$ is given by
the $\sharp$-conjugate of (\ref{eq:boltz}).

The value of $\tilde{\gamma}$ obtained from (\ref{eq:sigma11(1)}) is
not zero although, unlike $\alpha$, $\bar{\alpha}$, $\gamma$ and
$\bar{\gamma}$ it involves off-shell processes.  As discussed in the
last section, this implies that the thermal masses of particles and
antiparticles are different when the chemical potential is nonzero,
the difference being comparable in magnitude to the decay widths
$(\gamma/2 \pm {\rm Im}\Omega)$ of the quasiparticle modes. We are
nevertheless entitled to set $\tilde{\gamma}=0$. This simply means
that the difference in thermal masses is treated purely
perturbatively, rather that being resummed into the unperturbed
propagator. As it happens, the contributions involving
$\tilde{\gamma}$ to both (\ref{eq:Omegaeqn}) and (\ref{eq:Neqn}) are
of order $\lambda^4$. Thus, in low order calculations, it would be
consistent to retain a nonzero value of $\tilde{\gamma}$ in
(\ref{eq:hsoln}), while setting it to zero in the subsidiary
calculations needed to determine $\Omega(t)$ and $N(t)$. The
approximations needed to obtain the Boltzmann equation in any case
entail setting $\tilde{\gamma}=0$.
%%%%%%%%%%%%%%%%%%%%%%%%%%%%%%%%%%%%%%%%%%%%%%%
%% SECTION 6
%%%%%%%%%%%%%%%%%%%%%%%%%%%%%%%%%%%%%%%%%%%%%%%
\section{SUMMARY}
We have extended to the case of complex scalar fields a formulation of
non-equilibrium perturbation theory which partially resums the
dissipative effects of loop diagrams into the unperturbed propagator.
As a result of this resummation, low-order calculations using our
modified Feynman rules reflect the evolving state of the
non-equilibrium system through terms in the propagator which can
roughly be interpreted in terms of quasiparticle occupation numbers.
As in earlier work, we find that the evolution equation for these
occupation numbers can be cast approximately in the form of a
Boltzmann equation.  Kinetic equations of a similar form can, of
course, be derived by other methods (see, for example
\cite{calzetta2,boyan5}) if one sets about finding the rate of change
of occupation numbers defined in some appropriate manner. In an
interacting theory, however, the definition of single-particle modes
is generally ambiguous.  We, on the other hand, have set about
calculating unambiguously defined Green's functions in an optimal
manner.  For us, the quasiparticle modes and their occupation numbers
simply provide a convenient way of thinking about quantities which
arise naturally in the course of these calculations, and the Boltzmann
equation is a low-order approximation to a rather more complicated
second-order evolution equation.

The novel aspects of the work reported here concern the effects of a
non-zero chemical potential. To the extent that the quasiparticle
picture is valid, we find that not only the occupation numbers but
also the decay widths and thermal masses of the quasiparticles are
different for particle and antiparticle modes. In particular, the
difference in thermal masses is a two-loop effect (in the $\lambda
(\phi^*\phi)^2$ theory), which is not readily apparent in the usual
perturbative treatment (see e.g. \cite{kapusta}), although it should
be derivable from a two-loop calculation of the self-energy, even in
equilibrium.
%%%%%%%%%%%%%%%%%%%%%%%%%%%%%%%%%%%%%%%%%%%%%%%
\acknowledgments
D.B. would like to thank the University of Leeds for financial support
in the form of the William Wright Smith research scholarship.
%%%%%%%%%%%%%%%%%%%%%%%%%%%%%%%%%%%%%%%%%%%%%%%
% APPENDIX
%%%%%%%%%%%%%%%%%%%%%%%%%%%%%%%%%%%%%%%%%%%%%%%
\appendix
\section{}
Boundary conditions on the nine propagators $g_{ab}$ arise from two
sources. First, boundary conditions on the fields (\ref{eq:bcsonphi})
together with $\phi_1({\bf x},t_f)=\phi_2({\bf x},t_f)$, $\phi_2({\bf
x},t_0) = \phi_3({\bf x},0)$ and the complex conjugates of these lead
to
%%%%%%%%%%%%%%%%%%%%%%%%%%%%%%%%%%%%%%%%%%%%%%%
\begin{equation}
\begin{array}{rcl}
g_{1a}(t_f,t')&=&g_{2a}(t_f,t') \\
g_{2a}(t_0,t')&=&g_{3a}(0,t') \\
g_{1a}(t_0,t')&=&e^{\beta\mu}g_{3a}(0,t')
\end{array}\quad
\begin{array}{rcl}
g_{a1}(t,t_f)&=&g_{a2}(t,t_f)\\
g_{a2}(t,t_0)&=&g_{a3}(t,0)\\
g_{a1}(t,t_0)&=&e^{-\beta\mu} g_{a3}(t,0)
\end{array}
\end{equation}
%%%%%%%%%%%%%%%%%%%%%%%%%%%%%%%%%%%%%%%%%%%%%%%
for $(a=1,2,3)$. Second, the evaluation of the Gaussian path integral
leading to (\ref{eq:splitgen}) is performed, as usual, by completing
the square and this involves integrations by parts. The requirement
that boundary terms arising from these integrations by parts cancel
leads to
%%%%%%%%%%%%%%%%%%%%%%%%%%%%%%%%%%%%%%%%%%%%%%%
\begin{eqnarray}
\left. \partial_t g_{1a}(t,t')\right\vert_{t=t_f} &=&\left. \partial_t
g_{2a}(t,t') \right\vert_{t=t_f} \\
\lefteqn{
\left. \partial_t g_{2a}(t,t')\right\vert_{t=t_0}+\frac{1}{2}(\gamma +
\bar{\gamma})g_{1a}(t_0,t')-\frac{1}{2}(\bar{\gamma}+i\tilde{\gamma})
g_{2a}(t_0,t')}
\qquad\qquad\qquad\qquad\qquad\qquad && \nonumber \\
&=&\left. i\partial_{\tau}g_{3a}(\tau,t')\right\vert_{\tau=0}
+ \frac{\delta}{2}g_{3a}(0,t') \\
\lefteqn{
\left.\partial_t g_{1a}(t,t')\right\vert_{t=t_0} + \frac{1}{2}(\gamma
-\bar{\gamma})g_{2a}(t_0,t')+\frac{1}{2}(\bar{\gamma}-i\tilde{\gamma})
g_{1a}(t_0,t')} \qquad\qquad\qquad\qquad\qquad\qquad && \nonumber \\
&=&e^{\beta\mu}\left[\left.i\partial_{\tau}g_{3a}(\tau,t')\right\vert_
{\tau=\beta} + \frac{\delta}{2}g_{3a}(\beta,t')\right]\label {eq:dbcs}
\end{eqnarray}
%%%%%%%%%%%%%%%%%%%%%%%%%%%%%%%%%%%%%%%%%%%%%%%
for $(a=1,2,3)$, together with the $\sharp$-conjugates of these. The
coefficient $\delta$ corresponds to an additional counterterm which,
as explained in detail in \cite{idl:jpa}, enables us to impose
$\dot{N}(t_0)=0$ as befits an initial equilibrium state. As in
\cite{idl:jpa}, we find $\delta=-\gamma$. Here, $\gamma$,
$\bar{\gamma}$ and $\tilde{\gamma}$ are all evaluated at $t_0$.

The real-time propagators satisfying these boundary conditions are
described in the text.  Of those involving imaginary times, three are
found to be
%%%%%%%%%%%%%%%%%%%%%%%%%%%%%%%%%%%%%%%%%%%%%%%
\begin{equation}
g_{33}(\tau,\tau')=\frac{1}{2\omega}\left\{\left[n^- +\Theta(\tau -
\tau')\right]e^{-\omega(\tau-\tau')} + \left[n^++\Theta(\tau'-\tau)
\right]e^{\omega(\tau-\tau')}\right\}
\end{equation}
%%%%%%%%%%%%%%%%%%%%%%%%%%%%%%%%%%%%%%%%%%%%%%%
and
%%%%%%%%%%%%%%%%%%%%%%%%%%%%%%%%%%%%%%%%%%%%%%%
\begin{eqnarray}
g_{13}(t,\tau)&=&g_{23}(t,\tau)\nonumber\\
&=&\frac{e^{-\frac{1}{2}\int_{t_0}^t(\gamma(\bar{t})- i\tilde{\gamma}
(\bar{t}))d\bar{t}}}{4\sqrt{\Omega(t_0)\Omega(t)}}
\nonumber \\   &&\times
\left[(A_1e^{-\omega\tau}+A_2e^{\omega\tau})e^{-i\int_{t_0}^t\Omega(
\bar{t})d\bar{t}}
+(A_3e^{-\omega\tau}+A_4e^{\omega\tau})e^{i\int_{t_0}^t
\Omega(\bar{t})d\bar{t}}\right]\nonumber \\ &&
\end{eqnarray}
%%%%%%%%%%%%%%%%%%%%%%%%%%%%%%%%%%%%%%%%%%%%%%%
with
%%%%%%%%%%%%%%%%%%%%%%%%%%%%%%%%%%%%%%%%%%%%%%%
\begin{eqnarray}
A_1&=&(1+n^+)[N(t_0)-1-2n^-]/(1+n^++n^-)\\
A_2&=&n^-[N(t_0)+1+2n^+]/(1+n^++n^-)\\
A_3&=&(1+n^+)[N^{\sharp}(t_0)+1+2n^-]/(1+n^++n^-)\\
A_4&=&n^-[N^{\sharp}(t_0)-1-2n^+]/(1+n^++n^-)\ .
\end{eqnarray}
%%%%%%%%%%%%%%%%%%%%%%%%%%%%%%%%%%%%%%%%%%%%%%%
The others are given by
%%%%%%%%%%%%%%%%%%%%%%%%%%%%%%%%%%%%%%%%%%%%%%%
\begin{equation}
g_{31}(\tau,t) = g_{32}(\tau,t) = g_{13}^{* \sharp}(t,\tau)
\end{equation}
%%%%%%%%%%%%%%%%%%%%%%%%%%%%%%%%%%%%%%%%%%%%%%
Indeed, the whole propagator matrix has the property
%%%%%%%%%%%%%%%%%%%%%%%%%%%%%%%%%%%%%%%%%%%%%%%
\begin{equation}
g_{ab}^{* \sharp}(t,t') = g_{ba}(t',t)
\end{equation}
%%%%%%%%%%%%%%%%%%%%%%%%%%%%%%%%%%%%%%%%%%%%%%%
where $t$ and $t'$ denote real or imaginary times, depending on the
indices $a$ and $b$.
%%%%%%%%%%%%%%%%%%%%%%%%%%%%%%%%%%%%%%%%%%%%%%%
%  REFERENCES
%%%%%%%%%%%%%%%%%%%%%%%%%%%%%%%%%%%%%%%%%%%%%%%

%%%%%%%%%%%%%%%%%%%%%%%%%%%%%%%%%%%%%%%%%%%%%%%
%  FIGURE CAPTIONS
%%%%%%%%%%%%%%%%%%%%%%%%%%%%%%%%%%%%%%%%%%%%%%%
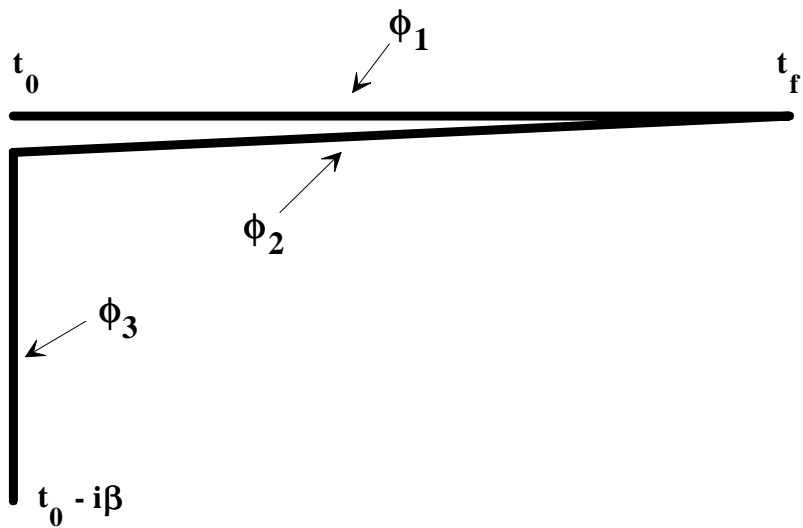
\begin{figure}
\caption{The closed time path in the complex time plane}
\label{Fig1}
\end{figure}
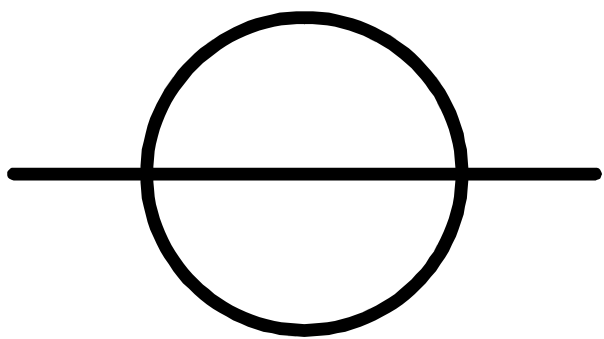
\begin{figure}
\caption{Lowest-order Feynman diagram contributing to the dissipative
counterterm}
\label{Fig2}
\end{figure}
%%%%%%%%%%%%%%%%%%%%%%%%%%%%%%%%%%%%%%%%%%%%%%%
\end{document}